\documentclass[showpacs,preprintnumbers,amsmath,amssymb]{revtex4}

\usepackage{graphicx}
\usepackage{array}

\begin{document}
\title{Crystal nucleation for a model of globular proteins}
\author{Andrey Shiryayev and James D. Gunton \\Department of
Physics, Lehigh University, Bethlehem, PA 18015}
\begin{abstract}
A continuum model of globular proteins proposed by Talanquer and
Oxtoby (J. Chem. Phys. \textbf{109}, 223  (1998)) is investigated
numerically, with particular emphasis on the region near the
metastable fluid-fluid coexistence curve. Classical nucleation
theory is shown to be invalid not only in the vicinity of the
metastable critical point but also close to the liquidus line. An
approximate analytic solution is also presented for the shape and
properties of the nucleating crystal droplet.
\end{abstract}
\maketitle

\section{Introduction}
In recent years there has been an enormous increase in the number
of proteins that can be isolated, due to the rapid advances in
biotechnology. However, the determination of the function of these
proteins has been slowed by the difficulty of determining their
crystal structure by standard X-ray crystallography.  A major
problem is that it is difficult to grow good quality protein
crystals.  Experiments have clearly shown that this
crystallization depends sensitively on the physical factors of the
initial solution of proteins. An important observation was made by
George and Wilson \cite{Wilson94}, who showed that x-ray quality
globular protein crystals only result when the second virial
coefficient, $B_2$, of the osmotic pressure of the protein in
solution lies within a narrow range. This corresponds to a rather
narrow temperature window. For large positive $B_2$,
crystallization does not occur on observable time scales, whereas
for large negative $B_2$, amorphous precipitation occurs.
Rosenbaum, Zamora and Zukoski then showed \cite{Zukoski96_1} the
crystallization of globular proteins could be explained as arising
from attractive interactions whose range is small compared with
the molecule's diameter (corresponding to the narrow window of
$B_2$. In this case the gas-fluid coexistence curve is in a
metastable region below the liquidus-solidus coexistence lines,
terminating in a metastable critical point. In a seminal study,
ten Wolde and Frenkel then showed \cite{Frenk97_1} in a simulation
of a modified Lennard-Jones model with short range attractions
that crystal nucleation is significantly enhanced in the vicinity
of the metastable critical point, due to a reduction in the
nucleation free energy barrier. Thus the conditions for optimal
crystallization correspond to being near the metastable critical
point.  The reduction in the nucleation barrier results from the
large density fluctuations that exist near the metastable critical
point, which also affect the structure of the critical nucleus.
Namely, the initial step toward formation of the critical droplet
is the creation of a dense, liquid-like droplet which then forms a
crystal nucleus in its interior. This process of first increasing
the density and then forming a crystalline order is the opposite
of what normally occurs in the liquid-solid phase transition at
high temperatures and close to the liquidus line. The authors also
showed that classical nucleation theory is invalid in the vicinity
of the critical point.
\\
\\
The idea that protein crystal nucleation could be enhanced by
critical density fluctuations was subsequently reinforced in a
density functional study by Talanquer and Oxtoby. They studied a
phase field model with two order parameters, the local density
$\rho(r)$ and a structural order parameter $m(r)$.  The
phenomenological free energy functional was assumed to have a van
der Waals fluid branch and a van der Waals-like solid branch.
Their numerical solution of the Euler-Lagrange equations for the
nucleating critical droplet and free energy barrier confirmed that
the nature of crystal nucleation changes qualitatively in the
vicinity of the metastable critical point. They also found that
the nucleation barrier is significantly lower in this region than
elsewhere in the phase diagram and that classical theory is
invalid.  Subsequently Sear studied a continuum model in which the
structural order parameter was omitted \cite{Sear01}. He carried
out a mean field theory near the metastable critical point and
showed that the liquid "tail" of the density profile of the
critical droplet is described by an Ornstein-Zernike like
exponential behavior, whose range is given by the fluid
correlation length. Since this length diverges as one approaches
the critical point, so does the critical size of the droplet and
the excess number of particles within it.  In a subsequent paper
Sear \cite{Sear02} extended his arguments via a scaling theory
using the critical exponents for the Ising universality class, so
as to be valid in the vicinity of the critical point. In
particular, the singular part of the free energy barrier
associated with the tail was shown to satisfy the same scaling as
the order parameter associated with the metastable critical point
(the density difference) and vanishes as one approaches $T_c$.
Sear did not calculate the shape of the droplet in the core and
interface region and hence was unable to calculate the value of
the barrier to nucleation.

 In this paper we extend the work of Talanquer and Oxtoby,
  to obtain a better understanding of homogeneous crystal
nucleation.  We present  additional numerical results for the
model as well as an approximate analytical description of the
critical droplet, including its crystal core and tail (the latter
is based on Sear's work \cite{Sear01}). From this we compute the
free-energy barrier to nucleation in the region between the
fluid-solid coexistence curve and the metastable fluid-fluid
coexistence curve.  As expected, this barrier is smallest in the
vicinity of the metastable critical point.  We show that classical
nucleation theory is incorrect in the region between the liquidus
and solid coexistence lines, except presumably very close to the
liquidus line.

The outline of the paper is the following.  In section 2 we
provide a brief review of the model.  In section 3 we extend the
numerical analysis of \cite{TalanquerOxtoby_98} for the nucleation
free energy barrier along various thermodynamic paths.  We also
calculate the profile of the critical droplet as a function of the
density and structural order parameter.  This yields the excess
number of molecules, N , and the number of molecules in the
crystalline phase, $N_c$ ,of the critical droplet. In the next
section we present a brief discussion of the range of validity of
classical nucleation theory and show that it is incorrect in the
entire metastable region of the phase diagram that we explore. We
also show results for the surface tension along different paths in
the phase diagram. In section 5 we present an approximate theory
for the shape of the critical droplet and from this obtain N and
$N_c$.  The theory accurately describes the critical droplet
except in the interface region, and is in qualitative agreement
with the numerical results for $N/N_c$. Section 6 contains a brief
conclusion.  Some details of the numerical analysis are given in
an Appendix.

\section{ Model} We use a model due to Talanquer and Oxtoby
\cite{TalanquerOxtoby_98} to describe globular protein
crystallization.  Their phase field model is based on the
following grand canonical free-energy functional:
\begin{equation}
\Omega\left[\rho, m\right] = \int d\vec{r} \left[f(\rho, m) -
\mu\rho +\frac{1}{2}K_{\rho}(\nabla\rho)^2
+\frac{1}{2}K_{m}\rho_{s}^2(\nabla m)^2 \right]
\label{Omega_FreeEnergy}
\end{equation}
The free energy depends on two order parameters: the (conserved)
local density $\rho(\textbf{r},t)$  and a (non-conserved) local
structural order parameter that shows whether the system is in a
solid or fluid phase $m(\textbf{r},t)$. Here $f(\rho,m)$ is the
Helmholtz free-energy density and $\mu$ is the chemical potential.
Talanquer and Oxtoby \cite{TalanquerOxtoby_98} use the van der
Waals free energy density for the fluid branch:
\begin{equation}
f_{f}(\rho,m) = k_{B}T\rho\left[ \ln\rho-1-\ln(1-\rho b) \right] -
a\rho^2 + k_{b}T\alpha_{1}m^2 \label{FluidFreeEnergy}
\end{equation}
and a corresponding van der Waals free energy for the solid phase:
\begin{equation}
f_{s}(\rho,m) = k_{B}T\rho\left[ \ln\rho-1-\ln(1-\bar{\rho} b)
\right] - (a+a_{m}m_{s}^2)\rho^2 + k_{b}T(\alpha_{1}m^2+\alpha_2)
\label{SolidFreeEnergy}
\end{equation}
Here $\bar{\rho}$ plays a role of a weighted coarse grained
density $\bar{\rho}=\rho(1-\alpha_3 m(2-m))$ where the term in
square brackets implements a difference between the fluid and
solid close-packing limits. For the fluid close-packing limit,
$\rho b=1 $, while for the solid close-packing limit
$\bar{\rho}b=1$. The quantity $m_s(\rho)$ is the equilibrium value
of the order parameter in the solid phase and hence is a solution
of the equation $(\partial f_s / \partial m)_{m_{s}(\rho)}=0$.
Thus in the solid close-packing limit ($b\bar{\rho}=1$) $m_s = 1$.
The parameter $a_m$ in equation (\ref{SolidFreeEnergy}) has been
introduced in order to change the range of the attractive
interactions between molecules in the solid phase from that in the
liquid phase (as described by the parameter a in equation
(\ref{FluidFreeEnergy})).

The chemical potential in the solid and fluid phases is the first
derivative of the free energy with respect to density: $\mu = (
\partial f / \partial\rho)_T$. In order to get coexisting densities $\rho_{\alpha}$ and
$\rho_{\beta}$ we have to solve the equations $\mu(\rho_{\alpha})
= \mu(\rho_{\beta})$ and $\omega(\rho_{\alpha}) =
\omega(\rho_{\beta})$, where $\omega = f-\mu\rho$. Graphically
this gives the well-known "common tangent" rule for coexistence.
By repeating this for different temperatures we can obtain the
entire phase diagram.

In order to obtain a critical droplet profile we must solve the
Euler-Lagrange equations with appropriate boundary conditions:
\begin{equation}
\frac{\delta\Omega}{\delta\rho} = 0 \text{  }\text{  } \text{ and
} \text{  }\text{  } \frac{\delta\Omega}{\delta m} = 0 \notag
\end{equation}
i.e.,
\begin{eqnarray}
-K_\rho\nabla^2\rho + \frac{\partial f}{\partial\rho} - \mu = 0
\label{rho_saddle_equation} \\
-K_m\nabla^2m + \frac{\partial f}{\partial m} = 0 \notag
\end{eqnarray}
Using the solutions of (\ref{rho_saddle_equation}) for
$\overline{\rho}(\textbf{r})$ and $\overline{m}(\textbf{r})$ we
can obtain such properties of the inhomogeneous system as the free
energy barrier $\Delta\Omega =
\Omega\{\overline{\rho}(\textbf{r}), \overline{m}(\textbf{r})\} -
\Omega\{\rho_0, 0\}$ and the surface tension $\sigma = 1/2(
\int_0^\infty K_{\rho}\left|\nabla\rho\right|^2 dr+\int_0^\infty
K_m\left|\nabla m\right|^2 dr )$.

\section{Numerical Results}

In Appendix A we describe the numerical methods that were used to
solving equations (\ref{rho_saddle_equation}). Here we present
some results, extending the work of \cite{TalanquerOxtoby_98}, for
the following choice of parameters: $a=1$, $b=1$, $\alpha_1 =
0.25$, $\alpha_2 = 2$, $\alpha_3= 0.3$, $a_m = 1$. The phase
diagram for these values is shown in Figure 1, which shows in
particular a metastable fluid-fluid coexistence curve. The
existence of the metastable critical point affects the nucleation
and growth processes in the vicinity of this point.

\begin{figure}[btp]
\center
\includegraphics[width=10cm]{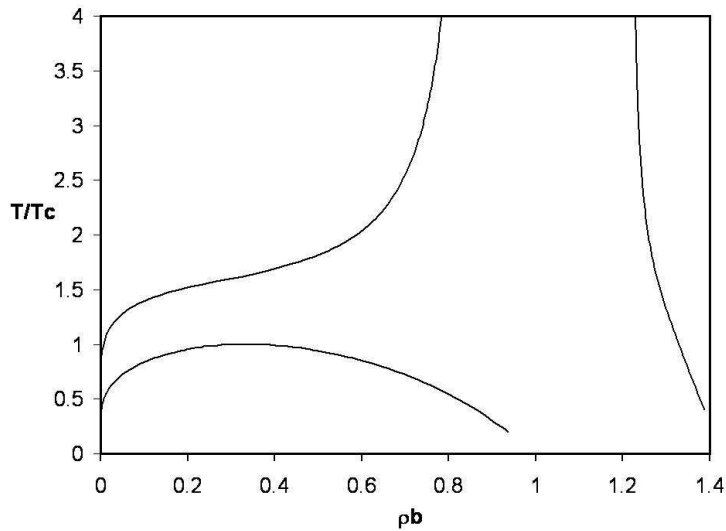}
 \caption{Phase diagram for the particles with the short-range
 attraction interactions. The fluid-fluid coexistence curve
 becomes metastable in this case.}
\end{figure}

The main quantity characterizing the nucleation process is a
nucleation rate $I$, given by
\begin{equation}
I=I_0 e^{-\Delta\Omega/kT}
\end{equation}
where $I_0$ is a prefactor given by the product of  dynamical and
statistical parts \cite{Langer69}. In order to understand how this
metastable critical point affects nucleation, we calculate the
free energy barrier for different thermodynamic paths. (Some
results for this barrier were presented in
\cite{TalanquerOxtoby_98}.) The barrier dependence on temperature
and density in the metastable region between the liquidus
(solubility curve) and solidus lines is shown in Figures 2 and 3.
Figure 2 shows that as the temperature decreases, the barrier also
decreases, while Figure 3 shows that as the density increases, the
barrier decreases. This is the normal behavior of a free energy
barrier, since at any point on the solubility curve the free
energy is infinite, decreases as one moves away from it and
vanishes at the solid coexistence curve.

\begin{figure}[btp]
\center
\includegraphics[width=11cm]{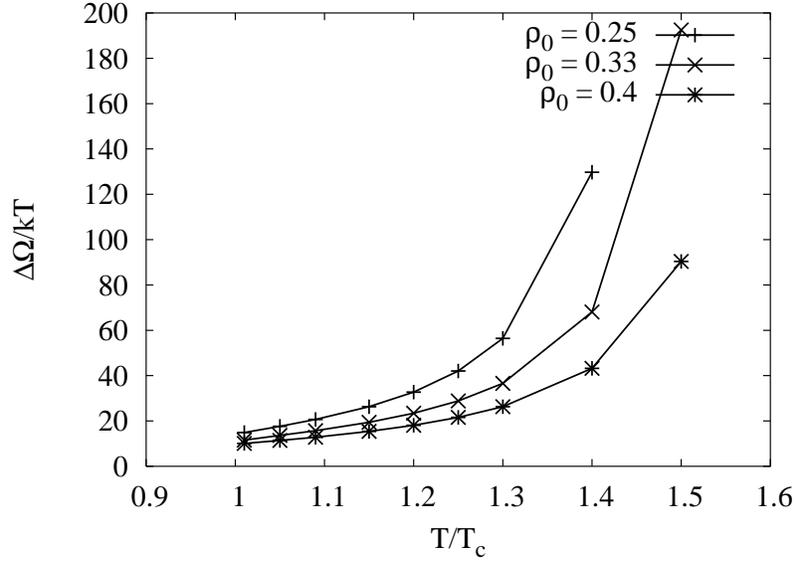}
 \caption{Dependence of the free energy barrier versus temperature at
 constant density.}
\end{figure}
\begin{figure}[btp]
\center
\includegraphics[width=11cm]{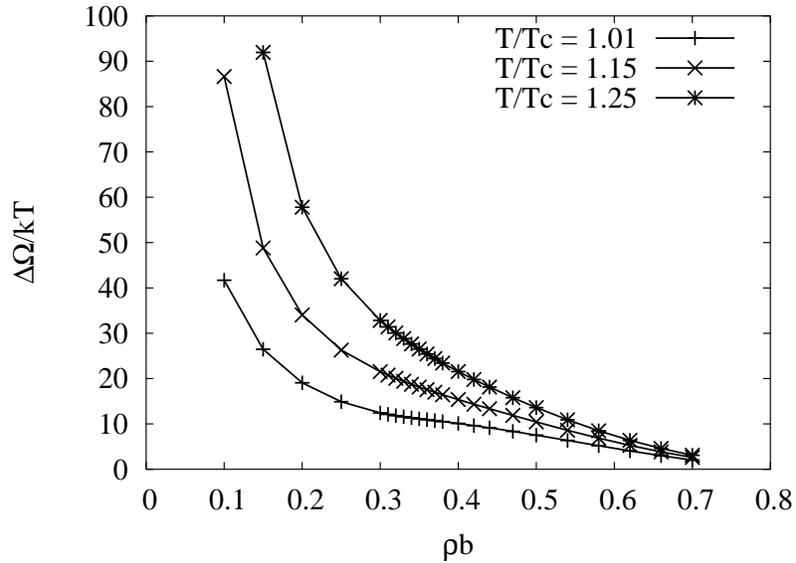}
 \caption{Dependence of the free energy barrier versus density at
 constant temperature.}
\end{figure}

However, the existence of the metastable fluid-fluid coexistence
curve changes the pathways of the constant free energy barrier and
constant supersaturation lines (Fig.4) as compared with the case
in which the coexistence curve is not metastable.  We can also see
that along the lines with constant supersaturation the free energy
barrier decreases as one approaches the critical point (Fig. 5),
which is consistent with \cite{TalanquerOxtoby_98}. Figure 5 shows
that the free energy barrier has a minimum near, but not at, the
critical point. The location of this minimum changes from above
the critical point for low supersaturation to below the critical
point for high supersaturation. It also can be seen that the
increase of the free energy barrier below the critical point
becomes very sharp for the cases where the constant
supersaturation lines intersect the fluid-fluid coexistence curve
\cite{Footnote_1}. In order to understand the behavior of the free
energy barrier along the constant supersaturation lines, we first
consider their shape (Fig. 4). At high temperatures these lines
are more vertical (constant density lines) and the free energy
barrier decreases with temperature, as shown in Fig. 2. Near  the
critical point the curves become almost horizontal (isothermal
lines) and the barrier starts to increase (Fig. 3). Thus the shape
of the constant supersaturation curve in some sense determines the
behavior of the free energy barrier. To understand this in more
detail, we note that the derivative of temperature with respect to
density at constant supersaturation is
\begin{equation}
\left( \frac{\partial T}{\partial\rho_0} \right)_{\Delta\mu} = -
\frac{\left(\frac{\partial\Delta\mu}{\partial\rho_0}\right)}
{\left(\frac{\partial\Delta\mu}{\partial T}\right)} =
-\left[\frac{f^{(f)}_{\rho\rho}}
{\left(\frac{\partial\Delta\mu}{\partial T}\right)} \right]_{\rho
= \rho_0} \label{dT_drho}
\end{equation}
where $f_{\rho\rho}$ is evaluated at the background fluid density.
This derivative vanishes at the critical point, so that the
constant supersaturation lines become horizontal in the vicinity
of the critical point. On the other hand, at large temperatures
the denominator vanishes, so that the constant supersaturation
lines become vertical. Thus we see that the presence of the
critical point changes the behavior of the free energy barrier
from decreasing as we lower the temperature far from the critical
point (the vertical part of the $\Delta\mu=const$ lines) to
increasing as we lower the temperature to its critical value (the
horizontal part of the $\Delta\mu=const$ lines). Therefore
somewhere in between there is a minimum of the free energy
barrier. Thus we can conclude that the free energy barrier is
relatively unaffected by the existence of the critical point along
paths of constant temperature or constant density , whereas it
plays a crucial role along paths of constant supersaturation

\begin{figure}[btp]
\center
\includegraphics[width=11cm]{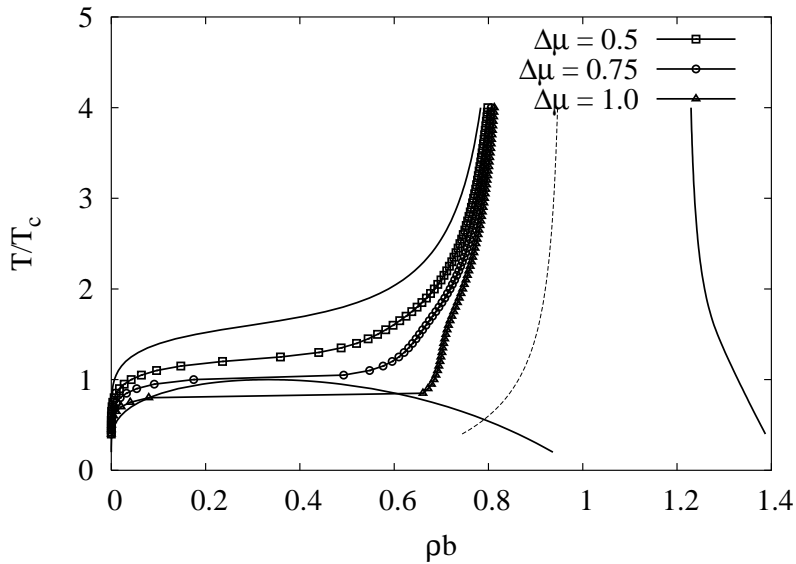}
 \caption{The solid lines on the phase diagrams show the paths of
 constant supersaturation. The dashed line shows the border between the background
fluid and background solid.  If the background density is to the
left of the dashed line, it corresponds to the fluid branch of the
free energy.  If it is to the right, it corresponds to the solid
branch.}
\end{figure}

\begin{figure}[btp]
\center
\includegraphics[width=11cm]{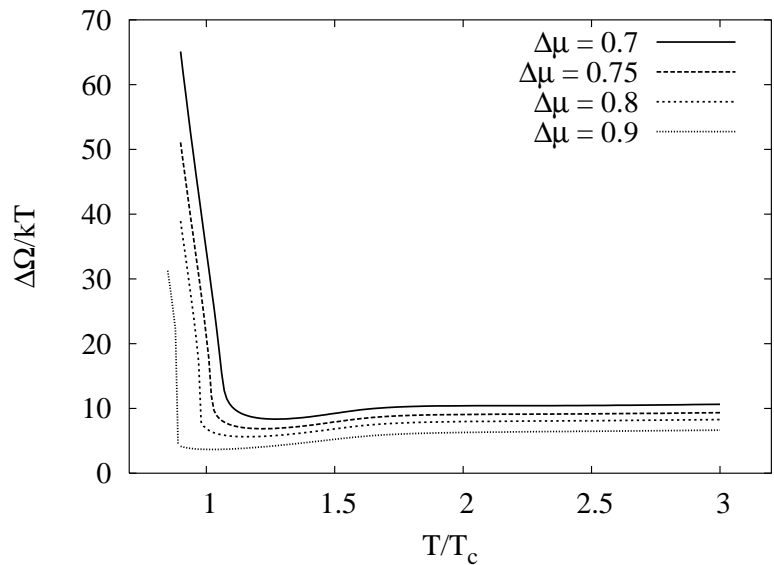}
 \caption{Dependence of the free energy barrier versus temperature at
 constant supersaturation.}
\end{figure}

One quantity of interest is the excess number of molecules,
defined as the number of molecules in the presence of the droplet
relative to the number of molecules in the spatially homogeneous
metastable state:
\begin{equation}
N = \int_V (\rho(r) - \rho_0)d\vec{r}
\end{equation}
We can calculate the number of molecules in the crystalline state
using $m(r)$, since this phase field shows whether a
 point
is in the solid or fluid state. Thus:
\begin{equation}
N_c = \int_V m(r)(\rho(r) - \rho_0)d\vec{r}
\end{equation}
Mean field theory yields a divergence in the excess number of
particles at the metastable critical point, whereas the number of
particles in the crystalline state remains finite. Figure 6 shows
$N_c$ versus $N$ for different supersaturations. We can check our
numerical results with the nucleation theorem \cite{NuclTheorem1,
NuclTheorem2}:
\begin{equation}
\frac{\partial\Delta\Omega}{\partial\Delta\mu} = -N
\end{equation}
Thus we can take a derivative of the free energy barrier with
respect to supersaturation and compare this with our results for
the dependence of the excess number of particles on the
supersaturation (Fig.7). We see that the free energy barrier
decreases rapidly near the critical point  as a function of the
supersaturation. This happens because the background fluid density
as a function of the supersaturation becomes flat in the vicinity
of the critical point.

\begin{figure}[btp]
\center
\includegraphics[width=11cm]{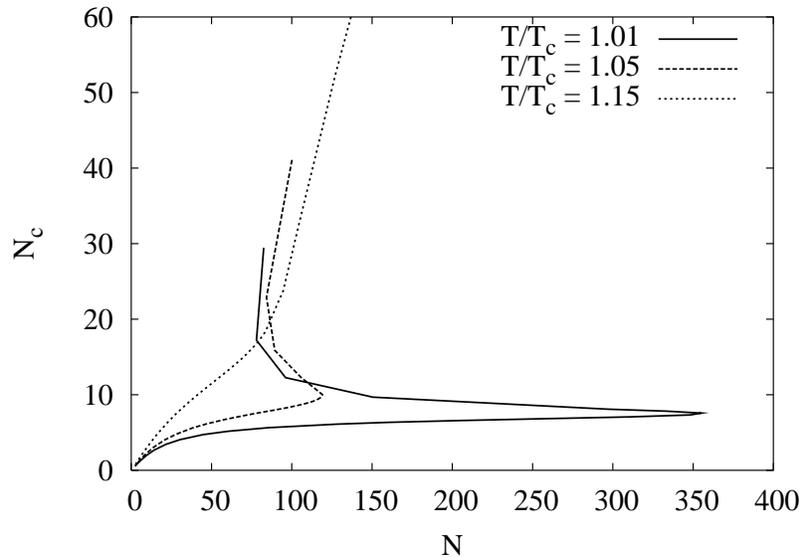}
 \caption{The number of crystalline particles versus
 excess number of particles for different temperatures.}
\end{figure}

\begin{figure}[btp]
\center
\includegraphics[width=11cm]{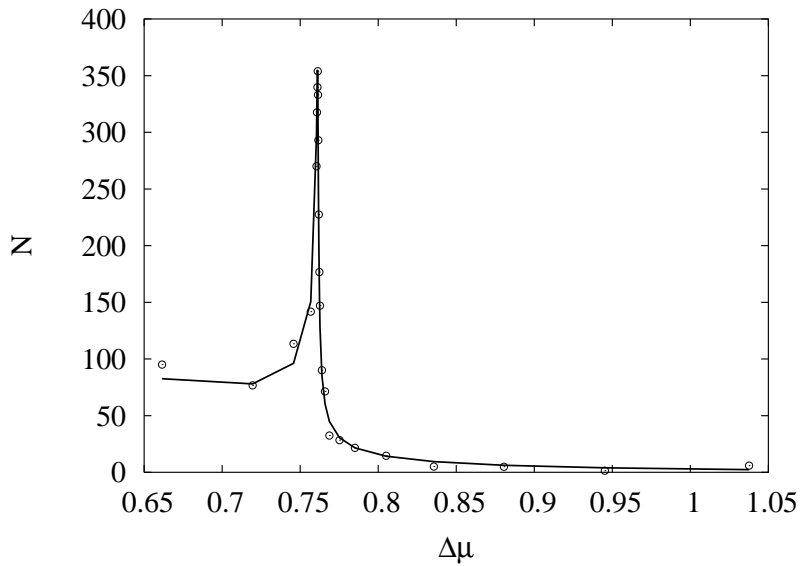}
 \caption{The excess number of particles versus
 supersaturation (solid line). The circles are obtained by taking
 the derivatives of the free energy barrier with respect to the
 supersaturation}
\end{figure}

\begin{figure}[btp]
\center
\includegraphics[width=11cm]{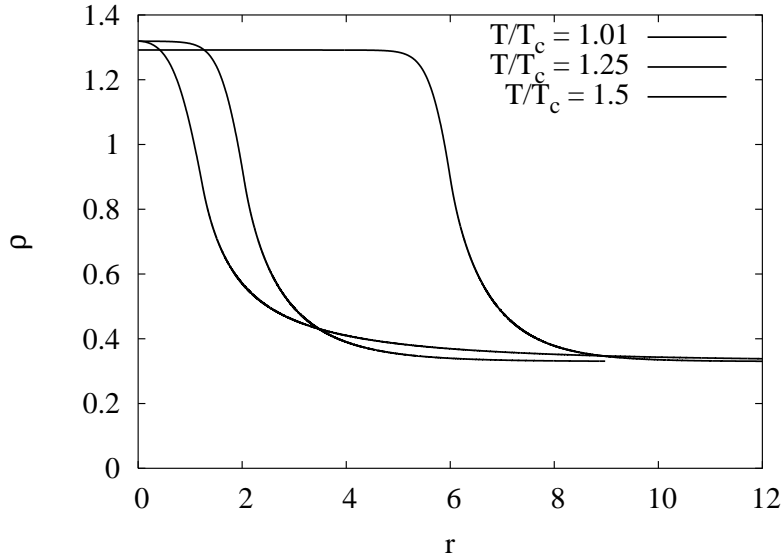}
 \caption{Density profiles for different temperatures. From left to
 right: T=1.01, T=1.25, T=1.5. The horizontal line shows the background
 fluid density.}
\end{figure}

\section{Breakdown of classical nucleation theory}

    Classical nucleation theory (CNT) assumes that the critical
droplet is large and that its interface is very sharp. It predicts
that the free energy barrier is
\begin{equation}
\Delta\Omega_{cl} =
\frac{16\pi\sigma^3}{3(\rho_s-\rho_0)^2\Delta\mu^2}
\end{equation}
where $\Delta\mu$ is $\mu - \mu_{coex}$.
 However, classical nucleation theory becomes increasingly
inaccurate as one approaches the metastable critical point
\cite{Frenk97_1, TalanquerOxtoby_98}. This is because the critical
droplet has a diffuse interface in this region, in contradiction
with the assumption of CNT.  Our goal in this section is to
compare CNT with our numerical results for the free energy barrier
and to show that CNT has at best a very limited region of
validity.  It is useful to divide the contributions to the free
energy barrier into two parts, i.e. the bulk and surface
contributions, and compare our numerical results with the
predictions of CNT for each of these separately.  First, define
\begin{equation}
\Delta\Omega_{bulk} = 4\pi\int \Delta\omega(r)r^2dr,
\end{equation}
\begin{equation}
\Delta\Omega_\sigma= 4\pi\int \frac{1}{2}\left[
K_\rho|\nabla\rho|^2 + K_m|\nabla m|^2\right]r^2dr
\end{equation}
so the free energy barrier can be written as
\begin{equation}
\Delta\Omega = \Delta\Omega_{bulk} + \Delta\Omega_\sigma.
\label{nonlinear_barrier}
\end{equation}
The classical approximation for this barrier is
\begin{equation}
\Delta\Omega_{cl} = \frac{4\pi R^3}{3}\Delta\omega_{cl} + 4\pi R^2\sigma
\label{classical_barrier}
\end{equation}
In the above $\Delta\omega(r) = [f(\rho(r),m(r)) - \mu\rho(r)] -
[f(\rho_0, 0) - \mu\rho_0]$ is the difference between the value of
the grand canonical free energy at the points $r$ and  infinity,
$\Delta\omega_{cl} = [f(\rho_s,m_s) - \mu\rho_s] - [f(\rho_0, 0) -
\mu\rho_0]$ and $\sigma$ is the surface tension. This division
also allows us to determine which of these contributions to the
free energy barrier is more important in the breakdown of  CNT. We
first note that if  CNT is correct at some point ($\rho$, $T$) of
the phase diagram, then the first terms of
(\ref{nonlinear_barrier}) and (\ref{classical_barrier}) should be
equal; likewise, the second terms also should be equal. Therefore
the values of the radii that result from comparison of the first
terms, i.e. $R_1 =
(3\Delta\Omega_{bulk}/|\Delta\omega_{cl}|)^{1/3}$, as well as the
second terms, $R_2 = (\Delta\Omega_{\sigma}/\sigma)^{1/2}$ should
be equal to each other and to the classical radius of the droplet,
$R_{cl} = 2\sigma/|\Delta\omega_{cl}|$.

Using this observation, we have checked the accuracy of CNT close
to the liquidus line at  low temperatures, as well as close to the
critical point. The point close to the critical point is $T/T_c =
1.01$ and $\rho b = 0.33$, and those close to the liquidus line
are $T/T_c = 1.5,\rho b = 0.33$ and  $T/T_c = 1.4, \rho = 0.2$.
For all these points CNT is incorrect. (For reasons described at
the end of the Appendix we have been unable to explore the region
very close to the liquidus line, where CNT presumably becomes
correct.) Figure 8 shows the density profiles for these cases,
while Table $1$ shows the values of the corresponding quantities,
$\Delta\Omega$, $\Delta\Omega_{cl}$, $R_1$, $R_2$, $R_{cl}$ and
the actual radius of the droplet $R_a$. We define the latter as
the radius of a spherical droplet which has an infinitely sharp
interface and the same solid density and number of particles as
our droplet. As can be seen from Table $1$, CNT underestimates the
nucleation barrier by a factor of ten as compared with the actual
value for the model. It also underestimates the critical droplet
size by a factor of two. As one would expect,  the ratio of the
actual and classical nucleation barriers decreases as we approach
the liquidus line. However, it is important to note that even in
the region which is 'close' to the liquidus line the difference
between the actual and classical values of the barrier is still
large.  It seems, therefore, that CNT is accurate only very close
to the liquidus line.

\setlength{\extrarowheight}{4pt}
\begin{table}
\begin{tabular}{|l|l|l|l|l|l|l|}
  \hline
 & $\Delta\Omega$ & $\Delta\Omega_{cl}$ & $R_1$ & $R_2$ & $R_{cl}$ & $R_{a}$ \\
  \hline
Close to liquidus line & 100.7 & 12.3 & 3.83 & 6.11 & 3.15 & 6.53 \\
  \hline
Close to critical point & 3.49 & 0.28 & 0.74 & 1.28 & 0.65 & 4.09 \\
  \hline
\end{tabular}
\caption{Values for comparison of our results with the classical
nucleation theory. The points close to the critical point and
liquidus line are $T/T_c = 1.01$ and $\rho b = 0.33$, and  $T/T_c
= 1.5$ and $\rho b = 0.33$, respectively.}
\end{table}

Next we try to determine how CNT breaks down. From Table 1 we see
that the radius $R_2$ extracted from the surface contribution to
the free energy barrier is closer to the actual value of the
droplet radius than the radius $R_1$ extracted from the bulk part.
In order to understand why $R_1$ is significantly smaller than the
actual radius, we examine the dependence of  $\Delta\omega$ on $r$
(Fig. 9). At the solid core  $\Delta\omega$ is negative and equal
to the classical $\Delta\omega_{cl} = \omega_{solid} -
\omega_{fluid}$ (dashed line in Fig. 9). As we reach the interface
region of the droplet, $\Delta\omega$ starts to increase, becomes
positive and reaches its maximum at the inflection point of the
density profile \cite{Footnote_2}. As $r$ increases beyond the
interface region, $\Delta\omega$ decays and approaches zero as we
reach the background value of the fluid density. In order to
obtain the bulk contribution to the barrier one must  multiply
$\Delta\omega$ by $r^2$ and then integrate over $r$. The
dependencies of $\Delta\omega(r)*r^2$ and the 'incomplete' barrier
$\Delta\Omega(r) = \int_0^r \Delta\omega(r)*r^2 dr$ on $r$ are
shown in Figures 10 and 11, respectively. The dashed lines in
these figures are $\Delta\omega_{cl}*r^2$ and $4\pi/3r^3
\Delta\omega_{cl}$. The dotted line in Figure 11 is the value of
the bulk contribution to the free energy barrier (which is equal
to the value of the incomplete barrier as $r\rightarrow \infty$).
The value of $R_1$ corresponds to the radius at which the bulk
contribution to the classical free energy is equal to the bulk
contribution of the actual free energy barrier (given by the
intersection of the dotted and dashed lines in Fig. 11).

\begin{figure}[btp]
\center
\includegraphics[width=11cm]{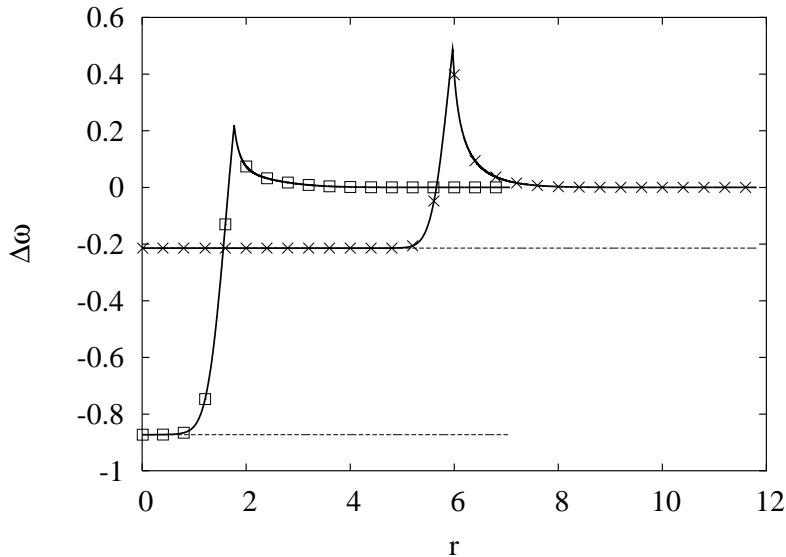}
 \caption{Grand canonical free energy density versus $r$ for two different
 points on the phase diagram.}
\end{figure}

\begin{figure}[btp]
\center
\includegraphics[width=11cm]{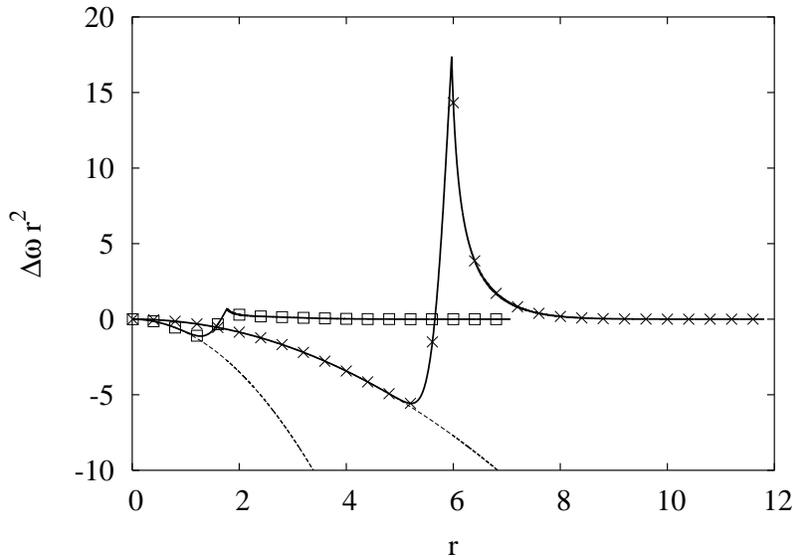}
 \caption{Grand canonical free energy density times $r^2$ (integrand)
 versus $r$}
\end{figure}

\begin{figure}[btp]
\center
\includegraphics[width=10cm]{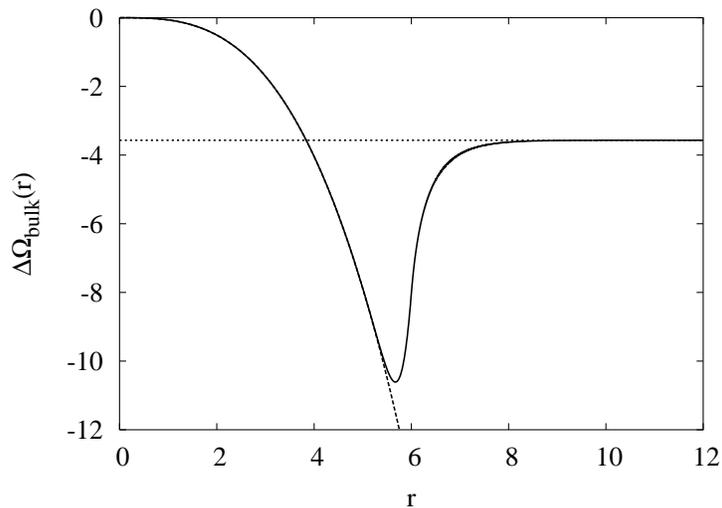}
 \caption{'Incomplete' bulk contribution to the free
 energy barrier versus $r$. Dashed line - dependence
 of the bulk contribution to the classical
    free energy
   barrier on the classical radius of the droplet. Dotted horizontal
   line is the bulk contribution of the free energy barrier.}
\end{figure}

Therefore, one can see that since the interface of the droplet is
not sufficiently sharp, the radius $R_1$, and therefore the
classical radius $R_{cl}$ , is significantly different from the
actual radius, $R_a$, of the droplet.  This is responsible for the
large difference between the classical and nonclassical values of
the free energy barrier. The main contribution to the interface is
its "tail", which is given by the correlation length and hence
becomes very large close to the critical point. At temperatures
close to the critical temperature the system should be in the very
narrow region near the fluid coexistence curve in order for the
radius of the droplet to be much larger than the interface
thickness, as given by the correlation length.

\section{Theoretical analysis}

\subsection{Approximations for the critical droplet.}
Equations (\ref{rho_saddle_equation}) with appropriate boundary
conditions define the saddle point of the free energy landscape in
functional space. The boundary conditions are a) that the density
and structural order parameter at infinity are given by the
metastable values and b) their spatial derivatives are zero at the
origin, i.e. $\rho(\infty) = \rho_0 \notag$, $m(\infty) = 0$,
$d\rho/dr(0) = 0$ and $dm/dr(0) = 0$. There is a trivial,
spatially uniform solution, satisfying these boundary conditions,
given by $\rho(r) = \rho_0$, $m(r)=0$. This uniform solution
corresponds to the supercooled liquid, which is a metastable
"point" in functional space. The critical droplet is a spatially
inhomogeneous, unstable
 solution satisfying the same boundary conditions, which we find by first
 solving
 the Euler-Lagrange equations around the droplet core ($r=0$).
 We assume that at the
origin the density has a value close to $\rho_s$ and the
structural order parameter has a value close to $m_s$.
\cite{Footnote_3} (In general the values of the density and
structural order parameter at the droplet center are not equal to
their equilibrium values, in contrast to the assumption of
classical nucleation theory.) First, we represent the solution in
the following form:
\begin{eqnarray}
\rho(r) = \rho_s + \frac{\chi_{\rho}(r)}{r} \label{rho_solution}\\
m(r) = m_s + \frac{\chi_m(r)}{r} \label{m_solution}
\end{eqnarray}
We substitute (\ref{rho_solution}) and (\ref{m_solution}) into
(\ref{rho_saddle_equation}), rewrite these equations in terms of
the $\chi$'s and then linearize them to obtain
\begin{eqnarray}
 -K_{\rho}\frac{d^2\chi_{\rho}}{dr^2} + f_{\rho\rho}\chi_{\rho} + f_{\rho m}\chi_m =
 0 \label{chirho_saddle_equation}
 \\
-K_{m}\frac{d^2\chi_m}{dr^2} + f_{m\rho}\chi_{\rho} +f_{m m}\chi_m
= 0 \label{chim_saddle_equation}
\end{eqnarray}
Here $f_{\rho\rho}$, $f_{\rho m}$ and $f_{m}$ are the second
derivatives of f  at  $\rho_s$ and $m_s$. Because this is an
approximation near the droplet core, we obviously cannot use the
boundary conditions at infinity. So, in addition to the zero
derivatives at $r=0$, we add two arbitrary boundary conditions at
the origin. The boundary conditions for the $\chi$'s therefore
become the following: $\chi_{\rho}(0) = 0$, $\chi_m(0) = 0$,
$\chi_{\rho}'(0) = \Delta\rho$ and $\chi_m'(0) = \Delta m$, where
$\Delta\rho$ and $\Delta m$ are the (unknown) shifts in the values
of density and order parameter from $\rho_s$ and $m_s$. The values
of $\Delta\rho$ and $\Delta m$ cannot be equal to zero, because in
this case the solution is spatially uniform (see Appendix A).
Solving  equations (17) and (18), we obtain

\begin{equation}
\chi_{\rho} = A \left( \frac{sinh(q_2r)}{q_2} -
\frac{sinh(q_1r)}{q_1} \right) + \frac{\Delta\rho}{q_2^2-q_1^2}
\left( \frac{q_2^2}{q_1}sinh(q_1r) - \frac{q_1^2}{q_2}sinh(q_2r)
\right) \label{chirho_complete_solution}
\end{equation}
and
\begin{equation}
\chi_m = B \left( \frac{sinh(q_2r)}{q_2} - \frac{sinh(q_1r)}{q_1}
\right) + \frac{\Delta m}{q_2^2-q_1^2} \left(
\frac{q_2^2}{q_1}sinh(q_1r) - \frac{q_1^2}{q_2}sinh(q_2r) \right)
\label{chim_complete_solution}
\end{equation}
where
\begin{eqnarray}
A = \frac{f_{\rho\rho}\Delta\rho + f_{\rho m}\Delta m}
{K_{\rho}(q_2^2 - q_1^2)} \notag \text{,  } B =
\frac{f_{m\rho}\Delta\rho + f_{m m}\Delta m} {K_{m}(q_2^2 -
q_1^2)} \notag \\
q^2 = \frac{1}{2}\left[ \left( \frac{f_{\rho\rho}}{K_{\rho}} +
\frac{f_{mm}}{K_m} \right) \pm \sqrt{ \left(
\frac{f_{\rho\rho}}{K_{\rho}} - \frac{f_{mm}}{K_m} \right)^2 +
\frac{4f_{\rho m}^2}{K_{\rho}K_m} } \right]
\end{eqnarray}
Thus we obtain the solution at the center of the droplet in terms
of two arbitrary parameters, $\Delta\rho$ and $\Delta m$. In
principle these parameters could be obtained from the exact
solution by matching this form to the boundary conditions at
infinity.

Next we find the solution for the tail of the droplet
\cite{Sear01} . As $r\rightarrow \infty$ we have metastable
boundary conditions, so we choose for the solution small
deviations from the metastable values:
\begin{eqnarray}
\rho(r) = \rho_0 + \frac{\delta_{\rho}}{r}e^{-q_{\rho}r} \notag  \\
m(r) = \frac{\delta_m}{r}e^{-q_m r}
\label{infinity_approx_solution}
\end{eqnarray}
where $\delta_{\rho}$ and $\delta_m$ are arbitrary parameters and
$q_{\rho} = 1/\xi^{(f)}_{\rho} =
(K_{\rho}/f^{(f)}_{\rho\rho})^{1/2}$ and $q_m = 1/\xi^{(f)}_m =
(K_m/f^{(f)}_{m m})^{1/2}$. The tail solution for $m$ is in fact
exact. We now have solutions in both the core and tail regions,
but not in the interface region. Figure 12 shows the comparison of
the numerically obtained density profile at $T/T_c = 1.2$ and
$\rho_0 = 0.33$ with our approximations for the tail and the core.
The tail approximation is rather good, but the core approximation
becomes incorrect as one approaches the interface region. For
comparison purposes, we used the  values of $\Delta\rho$, $\Delta
m$, $\delta\rho$ and $\delta m$ obtained numerically.

\begin{figure}[btp]
\center
\includegraphics[width=11cm]{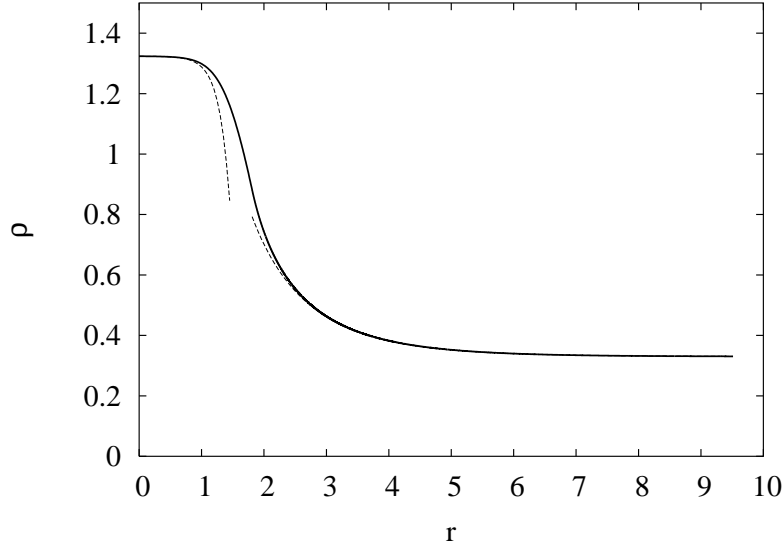}
 \caption{Comparison of the numerically obtained density profile
   (at $T/T_c = 1.2$ and $\rho_0 = 0.33$) with the tail and core
   approximations (dashed lines). }
\end{figure}

\subsection{Approximation for $N_c/N$.}

 $N$ and $N_c$ can each be considered as the sum of three terms: the
number in the core, the tail and  the interface. We can calculate
the contributions to each of these from the core and tail, using
the solutions for these two regions from the above section. Hence
we can write

\begin{equation}
N^{(core)} = \int_0^{R_{core}}(\rho_s + \frac{\chi_{\rho}(r)}{r} -
\rho_0)r^2dr \label{Ncore}
\end{equation}
\begin{equation}
N^{(tail)} = \int_{R_{tail}}^{\infty}
\frac{\delta_{\rho}e^{-q_{\rho}r}}{r}r^2dr \label{Ntail}
\end{equation}
where $R_{core}$ is the maximum value of $r$ for which we can
still use the approximation near the center of the droplet, and
$R_{tail}$ is the minimum value of $r$ for which the tail
approximation can be considered as correct. The interface region
is $R_{core} < r < R_{tail}$. In the core region,
$\frac{\chi(r)}{r} << \rho_s$ so we can drop the second term in
(\ref{Ncore}). Integrating (\ref{Ncore}) and  (\ref{Ntail}) we
find
\begin{eqnarray}
N^{(core)} = (\rho_s - \rho_0)V^{(core)} \\
N^{(tail)} = \delta_{\rho}\xi^{(f)}_{\rho}(\xi^{(f)}_{\rho} +
R_{tail})e^{-R_{tail}/\xi^{(f)}_{\rho}}
\end{eqnarray}
and similarly
\begin{equation}
N_c^{(core)} = \int_0^{R_{core}}(m_s + \frac{\chi_m(r)}{r})
(\rho_s + \frac{\chi_{\rho}(r)}{r} - \rho_0)r^2dr \label{Nc_core}
\end{equation}
\begin{equation}
N_c^{(tail)} = \int_{R_{tail}}^{\infty}
\frac{\delta_{\rho}e^{-q_{\rho}r}}{r} \frac{\delta_m e^{-q_m
r}}{r} r^2dr \label{Nc_tail}
\end{equation}
which gives us
\begin{eqnarray}
N_c^{(core)} = m_s(\rho_s - \rho_0)V^{(core)} \\
N_c^{(tail)} = \delta_{\rho}\delta_m
\frac{\xi^{(f)}_{\rho}\xi^{(f)}_m} {\xi^{(f)}_{\rho} +
\xi^{(f)}_m} e^{-R_{tail}(1/\xi^{(f)}_{\rho}+1/\xi^{(f)}_m)}
\end{eqnarray}

Next, we assume we can neglect contributions from the interface
region, i.e.
\begin{equation}
\frac{N_c^{(core)}+N_c^{(tail)}}{N^{(core)}+N^{(tail)}} \approx
\frac{N_c}{N}  \label{NcN_approximation}
\end{equation}
In order to check  this approximation, we compare its results with
the numerical results. However, to do this we must discuss how we
choose to determine the droplet profile, in order to estimate the
ratio of $N_c/N$. \cite{Footnote_5}

We first note that the LHS of (\ref{NcN_approximation}) depends on
two parameters, $R_{core}$ and $R_{tail}$ (for simplicity we
didn't introduce different radii for $\rho$ and $m$).  $R_{core}$
depends on the amplitudes $\Delta\rho$ and $\Delta m$, while
$R_{tail}$ depends on the amplitudes $\delta_{\rho}$ and
$\delta_m$. Therefore the LHS of (\ref{NcN_approximation}) depends
on four parameters. If we knew the interface solution
analytically, we would be able to determine these four parameters
by matching these solutions. As the approximation we could match
the core and the tail solutions (so $R_{core} = R_{tail}$). This
is the easiest to do but the least accurate. By setting $R_{core}
= R_{tail} = R_{match}$ we must take into account that the core
solution is obtained for the solid branch and the tail solution is
found for the fluid branch of the free energy. Therefore
$R_{match}$ is a point where the solid branch of the free energy
intersects the fluid branch.

\begin{figure}[btp]
\center \includegraphics[width=11cm]{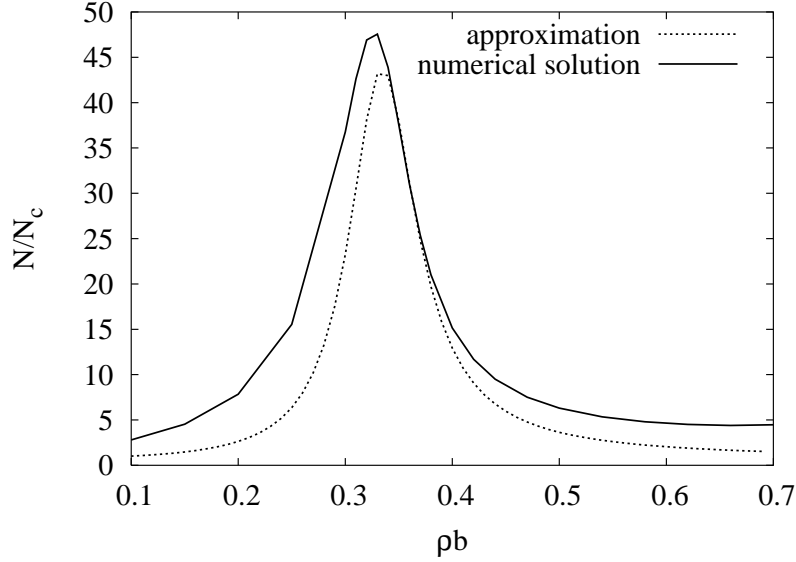}
 \caption{Approximation for the ratio of the excess number of
   particles and the number of crystalline particles
   versus background density for $T/T_c = 1.01$,
   obtained by matching core solution directly with tail solution.}
\end{figure}

Figure 13 shows the dependence of the ratio of the excess number
of particles and number of crystalline particles on the background
densities for the case in which we match the core solution
directly with the tail solution. As can be seen, this method gives
the best estimate for this ratio near the critical point, since
the tail plays a more important role in that region and the
approximation for the tail of $\rho$ is more accurate there.
However, the actual values of $N$ and $N_c$ are each
underestimated by a factor of ten .

As an alternative method we use the numerically determined
$\Delta\rho$ and $\Delta m$ to determine the core solution. Then
we match the tail solution with this core solution at the point of
intersection of the fluid and solid free energies. In this method
the derivatives of $\rho$ and $m$ with respect to $r$ are not
equal at the fitting point (Fig. 14). As  can be seen from Figures
15 and 16 this approximation is qualitatively good close to the
critical point, but becomes worse as $T/T_c$ becomes larger. We
still underestimate $N$ and $N_c$, but not by as large a factor as
in the above case.

\begin{figure}[btp]
\center
\includegraphics[width=10.5cm]{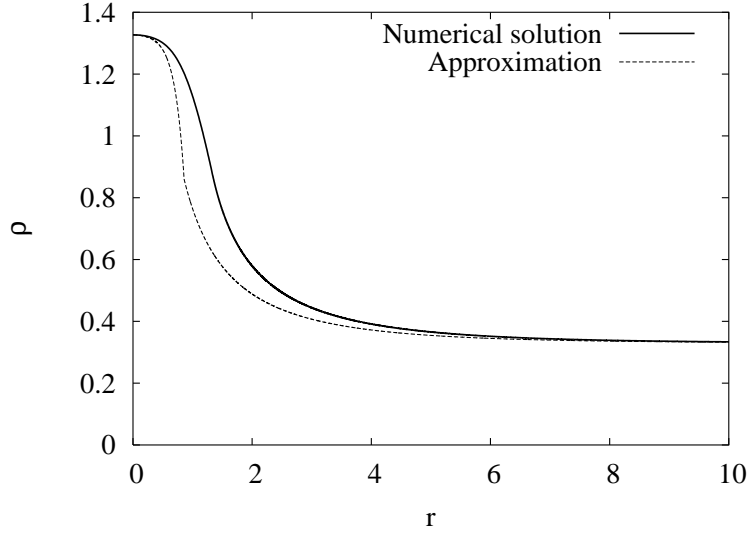}
 \caption{Comparison of the numerically obtained density profile
   (at $T/T_c = 1.2$ and $\rho_0 = 0.33$) with the solution obtained
   by matching core solution (with parameters $\Delta m$
   and $\Delta\rho$ obtained numerically) with tail solution,
   without matching the derivatives. }
\end{figure}

\begin{figure}[btp]
\center \includegraphics[width=10.5cm]{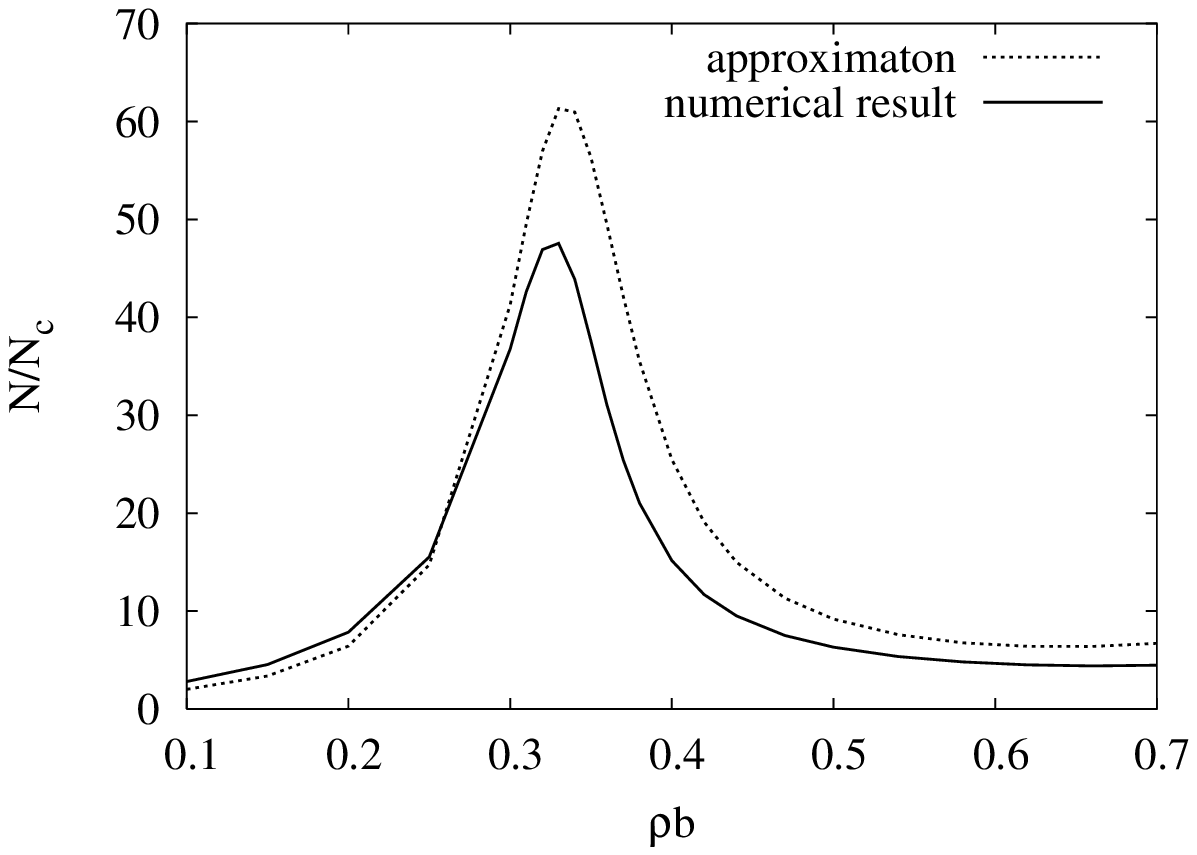}
 \caption{Approximation for the ratio of the excess number of
   particles and the number of crystalline particles
   versus background density for $T/T_c = 1.01$
   obtained by matching the core solution (with parameters $\Delta m$
   and $\Delta\rho$ obtained numerically) with the tail solution,
   without matching the derivatives.}
\end{figure}

\begin{figure}[btp]
\center \includegraphics[width=10.5cm]{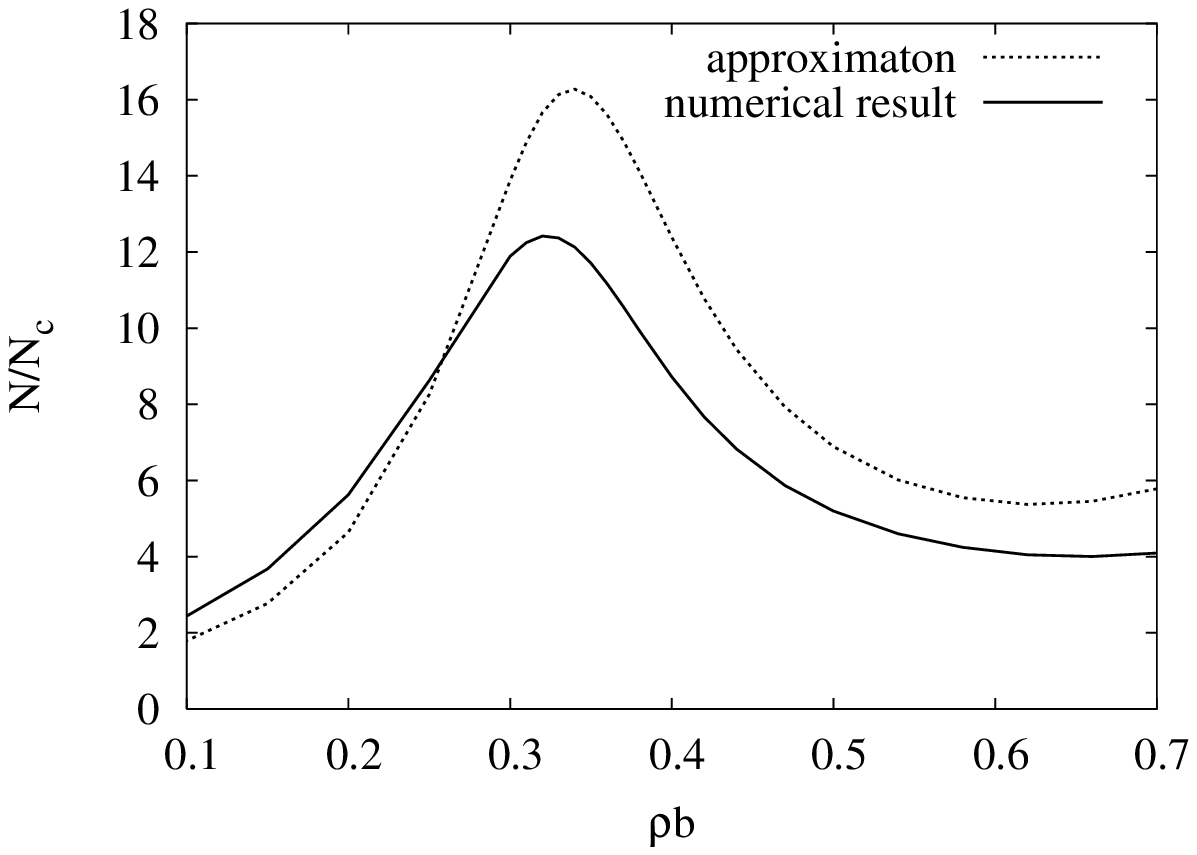}
 \caption{Approximation for the ratio of the excess number of
   particles and the number of crystalline particles
   versus background density for $T/T_c = 1.05$
   obtained by matching the core solution (with parameters $\Delta m$
   and $\Delta\rho$ obtained numerically) with the tail solution,
   without matching the derivatives.}
\end{figure}

However, close to the metastable critical point $N^{(tail)}$ (and
therefore $N$) diverges and becomes the leading term in the ratio
(\ref{NcN_approximation}),  while $N_c$ stays finite. Thus we can
obtain the asymptotic behavior of the ratio
(\ref{NcN_approximation}) close to the metastable critical point.
In this region we can rewrite (\ref{NcN_approximation}) as
\begin{equation}
\frac{N_c}{N} \approx \frac{N_c}{N^{(tail)}} \propto
\frac{1}{\left(\xi_\rho^{(f)}\right)^2} \propto f^{(f)}_{\rho\rho}
\label{NcN_criticalpoint}
\end{equation}
Therefore close to the metastable critical point this ratio
vanishes linearly with $\tau$ and quadratically with
$\Delta\rho_c$:
\begin{equation}
\left(\frac{N_c}{N}\right)_{metastable} \propto \left(\tau +
\frac{27}{8} (\Delta\rho_c)^2 \right) \label{NcN_criticalpoint1}
\end{equation}
where $\tau = (T-T_c)/T_c$ and $\Delta\rho_c = \rho_0 - \rho_c$.
Another asymptotic limit is the approach to the liquidus line. In
this case the radius of the solid core diverges and therefore
$N^{(core)}$ and $N_c^{(core)}$ become leading terms in (27) so at
the liquidus line the ratio becomes
\begin{equation}
\left(\frac{N_c}{N}\right)_{solubility} = m_s
\label{NcN_solubility}
\end{equation}

We can see that there are two cases where the number of molecules
in the critical droplet diverges, namely close to the critical
point and  close to the liquidus line, respectively. However, the
number of molecules in the solid core of the droplet diverges only
in the latter case. Thus close to the critical point most of the
molecules are in the tail of the droplet, due to the divergence of
the fluid correlation length,  as first pointed out by Sear
\cite{Sear01}. At the liquidus line most of the molecules are in
the solid core, whose radius diverges. Thus instead of one length
scale (the radius of the droplet) we have two: the radius of the
solid core and the fluid correlation length.  The solid core of
the globular protein crystal droplet can be observed
experimentally \cite{Vekilov99_1}, \cite{Vekilov00_1}, where it
has been shown that close to the metastable critical point the
core consists of only a few molecules( about 4-10) . However, the
existence of the long tail in the vicinity of the metastable
critical point has not been directly observed.

\section{Conclusion}
In this paper we have extended the numerical results
\cite{TalanquerOxtoby_98} to obtain a better understanding of
nucleation for their model. In particular, we have calculated the
density and structure order parameter profiles of the critical
nucleating droplet for different temperatures.  This solution of
the saddle point equations then yields the free energy barrier to
nucleation, which we have shown for a variety of different paths
in the phase diagram. We also show in detail how classical
nucleation theory is invalid everywhere in the region we studied,
except presumably very close to the liquidus line.  An approximate
theory for the shape and properties of the critical droplet is
developed that is in reasonable agreement with the numerical
results, although there is clearly a need for a better description
of the interface region that separates the core and tail of the
droplet. Finally, we note that experiments show the existence of a
gel state for globular proteins, in part of the metastable region
of the phase diagram \cite{Vekilov02}. As a consequence, our
discussion of the free energy barrier to nucleation for globular
proteins would only apply to the region of the phase diagram in
which there is no gelation, which is (roughly speaking) for
densities $\rho\leq \rho_c$. The model studied here does not
exhibit gelation.

\section{Acknowledgments:}  We wish to thank D. Oxtoby, V. Talanquer
and P. Vekilov for helpful discussions.  This work was supported
by a grant from the National Science Foundation,  DMR-0302598.

\section{Appendix A. Numerical solution of saddle point equations.}

Solving the equations (\ref{rho_saddle_equation}) is a nonlinear
two-boundary problem. One of the standard methods of solving this
type of equation is the shooting method. Because is it very
difficult to use the conditions at both boundaries simultaneously,
we implement the boundary conditions on one side, and choose the
number of free parameters equal to the number of boundary
conditions on the other side. Then we can solve the resulting
initial value problem. Depending on the final values (on the
second boundary), we can decide how to change these free
parameters. In our case we have two boundary conditions for $r=0$,
which we can use directly, and two boundary conditions at
$r=\infty$, which we cannot use directly. However, we can use
$\Delta\rho$ and $\Delta m$ as free parameters.

We still have several difficulties. First of all, the second
boundary is at infinity, so we can't explicitly get the values on
this boundary. We could use some large finite value instead of
infinity and avoid this problem, as long as the solution doesn't
diverge at infinity for an incorrect choice of the parameters
$\Delta\rho$ and $\Delta m$. Unfortunately it does diverge, so the
linear shooting method doesn't work in this simple way. The
reasons for the divergence are the error involved with the
numerical method and the absence of solutions for most choices of
$\Delta\rho$ and $\Delta m$. The error involved with the numerical
method increases as we approach a cusp in free energy landscape;
this is one of the reasons for the divergence. By varying the
parameters $\Delta\rho$ and $\Delta m$ we can push the beginning
of this divergence to larger distances, so we can obtain part of
the solution (including part of the tail) before the divergence
occurs. However, this gives large errors in estimating such
quantities as the free energy barrier, surface tension, etc. We
could decrease the numerical error of the shooting method in the
interface region by using Runge-Kutta methods with adaptive step
size, but if we desire reasonably small values of the errors, the
step size becomes too small for simulations \cite{Footnote_4}.

To avoid these difficulties we use a more  advanced shooting
technique - shooting with a fitting point. The fitting point was
chosen at $r=R_{core}$ where the solid and fluid branches of the
free energy intersect. The value of $R_{core}$ depends on the
parameters $\Delta\rho$ and $\Delta m$. Next, we know the exact
solution for the fluid part of $m(r)$
(\ref{infinity_approx_solution}), where $\delta_{\rho}$ is a
parameter. By matching the fluid and solid parts of $m(r)$ and
their derivatives at $r=R_{core}$ we determine $\Delta m$ for a
fixed values of $\Delta\rho$ and $\delta\rho$. The next step is to
obtain $\Delta\rho$ and $\delta\rho$ by matching the values and
derivatives of the fluid and solid parts of $\rho(r)$. Note that
the value of $\Delta m$ depends on the value of $\Delta\rho$, so
when we shoot with a particular value of $\Delta\rho$ we obtain a
value for $\Delta m$ each time. This shooting with a fitting point
minimizes the numerical error as compared with standard shotting
method.

There still is one remaining problem, even if we use shooting with
a fitting point. Namely, as we approach the fluid coexistence
curve, the values of $\Delta\rho$ and $\Delta m$ become very
small, so for shooting and fitting we need to vary them very
slightly. Thus it becomes beyond the numerical limit of precision
of the calculations to obtain a matching at the fitting point as
we change these parameters. This so far prevented us from
exploring the free energy barrier as one approaches the
coexistence curve and hence we have not been able to determine
where the classical theory becomes valid.


\begin{thebibliography}{10}
\bibitem{Footnote_1}
Because there is a discontinuity in those lines of constant
supersaturation
  which intersect the fluid-fluid coexistence curve, the free energy barrier
  has a corresponding discontinuity. this leads to the jumps in the dependence
  of the barrier on temperature (fig. 5 for $\delta \mu=0.8$ and $\delta
  \mu=0.9$.).
\bibitem{Footnote_2}
In our model the second derivative is not continuous, so we have a
cusp in the
  maximum.
\bibitem{Footnote_3}
This is true unless the critical droplet becomes very small, which
is the case
  for high supersaturations. we do not consider this case here.
\bibitem{Footnote_5}
It is obvious that the accuracy of the approximation
(\ref{NcN_approximation})
  depends on the accuracy of the core and tail solutions, as well as the
  relative contribution from the interface region. the maximum accuracy can be
  obtained by balancing between these two criteria. for example, if $r_{core}$
  is small, so the core solution is very accurate and $r_{tail}$ is large, so
  the tail solution is also very accurate, then the gap between $r_{core}$ and
  $r_{tail}$ becomes large and therefore the interface region plays an
  important role. and vice versa, if $r_{core}$ is close to $r_{tail}$ the
  interface region isn't that important, but the accuracy of the core and tail
  solutions can decrease dramatically.
\bibitem{Footnote_4}
This means that the step size $\delta r$ becomes so small that for
the computer
  $1 + \delta r$ becomes equal to $1$. in our simulation $\delta r$ is about
  $10^{-15}$.
\end{thebibliography}
\end{document}